\documentclass[conference]{IEEEtran}
\IEEEoverridecommandlockouts
\usepackage{amsmath,amsfonts}
\usepackage{amssymb,bbm,bm}
\usepackage{amsthm}
\usepackage{algorithmic}
\usepackage{algorithm}
\usepackage{array}
\usepackage[caption=false]{subfig}
\usepackage{textcomp}
\usepackage{stfloats}
\usepackage{url}
\usepackage{verbatim}
\usepackage{graphicx}
\usepackage[dvipsnames]{xcolor}
\usepackage{cite}
\usepackage{hyperref}
\usepackage{upgreek}
\usepackage{tikz}
\usepackage{pgfplots}
\usepackage{balance}
\usepgfplotslibrary{groupplots}
\pgfplotsset{compat=1.18}
\hypersetup{
    colorlinks=true,
    linkcolor=BrickRed,
    citecolor=blue,
    urlcolor=NavyBlue
}

\theoremstyle{definition}
\newtheorem{remark}{Remark}
\newcommand{\Prb}{\mathbbmss{P}}
\newcommand{\E}{\mathbbmss{E}}
\newcommand{\CN}{\mathcal{N}_{\mathbb{C}}}
\newcommand{\imag}{\mathrm{i}}

\begin{document}
\title{A Superposition Signaling Scheme for\\Integrated Sensing and Communication
\thanks{This work was supported by the European Research Council (ERC) under Grant 101116550.
The work of Giuseppe Caire was supported by the Gottfried Wilhelm Leibniz-Preis 2021 of the German Science Foundation (DFG).
}
}
\author{
\IEEEauthorblockN{
Gökhan~Yılmaz\IEEEauthorrefmark{1},
Hamdi~Joudeh\IEEEauthorrefmark{1}
and~Giuseppe~Caire\IEEEauthorrefmark{2}
}
\IEEEauthorblockA{
\IEEEauthorrefmark{1}%
Department of Electrical Engineering, 
Eindhoven University of Technology, 
The Netherlands
}
\IEEEauthorblockA{
\IEEEauthorrefmark{2}%
Faculty of Electrical Engineering and Computer Science,
Technical University of Berlin,
Germany
}
\IEEEauthorblockA{{\ttfamily\fontsize{9}{10}\selectfont
\IEEEauthorrefmark{1}\{g.yilmaz, h.joudeh\}@tue.nl,\;
\IEEEauthorrefmark{2}caire@tu-berlin.de}
}
}
\maketitle
\begin{abstract}
We investigate the capacity of an integrated sensing and communication system
operating with orthogonal frequency division multiplexing, where the integrated
sidelobe level of the transmit signal is adopted as an input cost.
The problem is reduced to the capacity of a complex additive white Gaussian noise channel under power and kurtosis constraints.
The sensing-optimal and communication-optimal operating points are characterized by circularly symmetric constant-modulus and complex Gaussian inputs, respectively, while the capacity-achieving input in between has discrete amplitude support with a concentric multi-ring geometry.
We propose the superposition of the two extremal inputs as a simple and tunable
signaling strategy, whose rate admits an exact expression via successive decoding.
Numerical results show that the proposed strategy attains a worst-case gap of approximately $0.043$ bits to the capacity.
\end{abstract}
\section{Introduction}
\label{sec:intro}
The integration of sensing and communication functionalities into a single wireless system, known as integrated sensing and communication (ISAC), is envisioned as a core feature of next-generation wireless networks \cite{FLiu2020}.
This has prompted a large body of work on the fundamental performance limits and dual-function waveform design of ISAC systems (see \cite{Ahmadipour2025,FLiu2025c} and references therein).
Such studies reveal a fundamental trade-off between sensing and communication: signals that are well suited for communication are generally suboptimal for sensing, and vice versa. 
A central problem in ISAC research is therefore the design of flexible signaling schemes that enable a tunable trade-off between the two functionalities.

In this paper, we study this question for ISAC systems operating with orthogonal frequency division multiplexing (OFDM). 
Our focus on OFDM is motivated by both practical and theoretical considerations. OFDM is the backbone of current wireless standards and is expected to remain central to future networks. 
Moreover, it seamlessly combines communication with delay-Doppler recovery \cite{Sturm2011}, and has recently been shown to enjoy favorable sensing properties in low-mobility scenarios, in terms of both the ambiguity function sidelobes \cite{FLiu2025a,Zhang2025} and the ranging Cramér-Rao bound \cite{FLiu2026}.

In the current work, we consider the delay-Doppler recovery sensing task and make the low-mobility assumption, which is relevant in, e.g., indoor sensing \cite{3GPP22837} and
cellular radar imaging \cite{Hu2026}.
To make the communication-sensing trade-off concrete, we must fix a sensing metric. We adopt the integrated sidelobe level (ISL), which originates in the radar signal processing literature \cite{Stoica2009a,Stoica2009b}, and measures the sidelobe power of the ambiguity function (we further adopt the common convention of normalizing the ISL by the main-lobe power \cite{Stoica2009b,Zhang2025}).
A larger ISL is associated with degraded sensing performance, e.g., reduced detection probability of the cell-averaging constant false alarm rate (CA-CFAR) detector \cite{Geiger2026}; and increased ranging error probability when on-grid ranging is cast as a multiple hypothesis testing problem \cite{Yilmaz2026b}.

Beyond its operational relevance, adopting the ISL leads to a clean cost-constrained capacity formulation, in which the mutual information of an additive white Gaussian noise (AWGN) channel is maximized subject to second- and fourth-moment (or kurtosis) constraints.
Solving this problem across ISL budgets traces out the communication-sensing
trade-off and reveals the structure of the signaling schemes that attain it.
This capacity problem was recently studied by Geiger \emph{et al.} \cite{Geiger2026}, who derive an upper bound by maximizing the output entropy under a set of induced moment constraints, and a lower bound via the entropy power inequality. 
They further use the kurtosis to guide the design of practical constellation shaping schemes (see also \cite{Du2024}, \cite{Shao2026}). 
The structure of the optimal (i.e., capacity-achieving) signaling strategy, however, was not addressed in \cite{Geiger2026}.
We address this gap in the present paper, and further propose an alternative simple signaling strategy which is shown numerically to approach capacity.
\subsubsection*{Contribution}
We first characterize the structure of the capacity-achieving input. The
communication-optimal and sensing-optimal endpoints are attained by the complex
Gaussian and the circularly symmetric constant-modulus distributions, respectively.
Between these endpoints, the fourth-moment constraint is always active, and by
specializing the analysis of Gursoy \emph{et al.} \cite{Gursoy2005}, the optimal
input can be shown to have discrete amplitude supported on a finite set, yielding a
concentric multi-ring geometry with continuous uniform phase. 
This provides a capacity-theoretic justification for amplitude-phase-shift keying (APSK) reported to perform well in the constellation shaping literature \cite{Geiger2026, Shao2026, Keshavarzchafjiri2026}.

Since the capacity-achieving input has no closed form, we propose a simple superposition signaling scheme in which the Gaussian and constant-modulus inputs are superimposed with a single power-allocation parameter, tuned to meet the designated ISL budget. 
The rate of this scheme admits an exact expression, obtained via successive decoding, which decomposes into the constant-modulus and Gaussian capacities at their respective effective signal-to-noise ratios (SNRs). The scheme is capacity-achieving at both communication- and sensing-optimal endpoints.
Numerically, the superposition scheme attains a worst-case gap of $\approx 0.043$ bits to the capacity across all considered SNRs and ISL budgets. This is despite the superposition input having continuous amplitude support, in contrast to the finite support of the optimal input.

\subsubsection*{Notation}
$\log(\cdot)$ denotes the natural logarithm.
$f_X(x)$ and $F_X(x)$ denote the probability density function (pdf) and the cumulative distribution function (cdf) of a random variable $X$, respectively.
If $X$ has a bounded pdf, then its differential entropy is denoted by $h(X)$.
The mutual information between two random variables $X$ and $Y$ is denoted by $I(X;Y)$.

\section{Problem Setting}
\label{sec:setting}
We consider an OFDM-based ISAC setting where a transmitter aims to communicate with a remote receiver and simultaneously generate backscattered signals to enable the sensing of targets in the environment.
The sensing task is the on-grid recovery of target delay-Doppler parameters \cite{Sturm2011}.
We consider a monostatic sensing setup where the sensor is colocated with the transmitter and has access to the transmit signal.

\subsection{Signal model}
The OFDM system employs $K$ subcarriers, and each transmission spans $N$ OFDM symbols.
A transmission block therefore comprises $K \times N$ channel symbols.
A cyclic prefix (CP) of length $L \leq K$ is appended to each OFDM symbol, chosen to cover the channel delay spread and to guarantee an unambiguous sensing range.
The channel symbol transmitted on the $k$-th subcarrier of the $n$-th OFDM symbol is denoted by $X_{k,n}$.
The CPs are removed at both the communication receiver and the radar sensor through standard OFDM processing, and hence do not appear in the signal model hereafter.

For simplicity, we focus on a frequency-flat communication channel that remains fixed throughout the transmission block.
The communication receiver observes
\begin{equation}
    Y_{k,n} = \alpha \, X_{k,n} + Z_{k,n} \label{eq:comm_signal}
\end{equation}
for $k=0,1,\dots,K-1$ and $n=0,1,\dots,N-1$, where $\alpha \in \mathbb{C}$ denotes the channel gain and $Z_{k,n} \sim \CN(0,\sigma^2)$ denotes zero-mean circularly symmetric Gaussian noise, independent across channel uses.
We assume the channel response is known to both the transmitter and receiver through an initial access phase conducted prior to transmission.
Hence, we set $\alpha = 1$ without loss of generality.

In a given transmission block of size $K \times N$, the transmitter conveys a message $W$ drawn uniformly at random from a message set $\mathcal{M} \triangleq \{1,2,\ldots,M\}$.
Given $W=w$, the transmitter sends the codeword $\bm{x}(w) \in \mathbb{C}^{K \times N}$, whose $(k,n)$-th entry is $x_{k,n}(w)$.
The codebook is the set of all codewords $\bm{\mathcal{C}} \triangleq \{ \bm{x}(1), \bm{x}(2), \dots, \bm{x}(M) \}$, and the corresponding communication rate is given by $\frac{1}{KN}\log M$ nats per complex channel use.
Each codeword is subject to a power constraint and to an ISL constraint, where the latter is imposed by the sensing requirements. Both constraints are specified next.
\subsection{Power and ISL constraints}
For ease of notation, we first define the empirical $\ell$-th moment of a codeword $\bm{x} \in \bm{\mathcal{C}}$ as
\begin{equation}
    \hat{\mu}_\ell(\bm{x}) \triangleq \frac{1}{KN} \sum_{k=0}^{K-1} \sum_{n=0}^{N-1} |x_{k,n}|^{\ell}.
\end{equation}
The power constraint on the codewords is
\begin{equation}
    \hat{\mu}_2(\bm{x}) \leq P
    \label{eq:power}
\end{equation}
for every $\bm{x} \in \bm{\mathcal{C}}$, where $P \geq 0$ is the power budget.

We now formalize the ISL cost constraint, which ensures that each codeword is also a \emph{good} probing signal, depending on the sensing requirement.
To this end, we define the discrete ambiguity function \cite{FLiu2025a, Zhang2025} of the codeword $\bm{x}$, evaluated at delay bin $i$ and Doppler bin $j$, as
\begin{equation}
    a_{i,j}(\bm{x} ) \triangleq \frac{1}{KN} \sum_{k=0}^{K-1} \sum_{n=0}^{N-1} |x_{k,n}|^2 \exp \! \left( \imag 2 \pi \left( \tfrac{ki}{K} - \tfrac{nj}{N} \right) \right) \label{eq:amb_func}
\end{equation}
where $(i,j)=(0,0)$ denotes the \emph{main-lobe} and the indices $(i,j) \neq (0,0)$ refer to the \emph{sidelobes}.
The normalization by $K \times N$ is for convenience, particularly since we later consider large $K$ and $N$.
For the discrete ambiguity function in \eqref{eq:amb_func} to be meaningful, a low-mobility regime is implicitly assumed, in which the maximal Doppler shift is small relative to the subcarrier spacing.
This enables modeling the phase shift induced by Doppler shifts as constant within one OFDM symbol (known as the \emph{fast-slow time} model in radar \cite{Zhang2025}).

For codeword $\bm{x}$, the ISL is defined as \cite{Stoica2009a,Stoica2009b,FLiu2025a,Zhang2025}
\begin{align}
\label{eq:ISL_def}
    \mathrm{ISL}(\bm{x}) & \triangleq  \left( \sum_{i=0}^{K-1} \sum_{j=0}^{N-1} |a_{i,j}(\bm{x})|^2 \right) - |a_{0,0}(\bm{x})|^2\\
    \label{eq:Parseval}
    & = \hat{\mu}_4(\bm{x}) - \big(\hat{\mu}_2(\bm{x})\big)^2
\end{align}
where \eqref{eq:Parseval} follows from Parseval's theorem.
The ISL has units of power-squared and scales with the transmit power. 
To obtain a cost that reflects the \emph{shape} of the ambiguity
function instead of its scaling, it is common to normalize the ISL by the main-lobe power $|a_{0,0}(\bm{x})|^2 = \big(\hat{\mu}_2(\bm{x})\big)^2$ \cite{Stoica2009b, Zhang2025}.
The normalized ISL constraint imposed on the codebook is hence given by
\begin{equation}
    \frac{\mathrm{ISL}(\bm{x})}{\big(\hat{\mu}_2(\bm{x})\big)^2}
    = \frac{\hat{\mu}_4(\bm{x})}{\big(\hat{\mu}_2(\bm{x})\big)^2} - 1 \leq S
    \label{eq:isl}
\end{equation}
for every $\bm{x} \in \bm{\mathcal{C}}$, where $S \geq 0$ is the (normalized) ISL budget.
Hence the ISL cost is governed by the empirical second and fourth moments of the
codeword, and \eqref{eq:isl} can equivalently be seen as a constraint on the \emph{empirical kurtosis}.
Note that a higher ISL budget is associated with degraded sensing performance and improved communication performance.

We are interested in the performance limits of such cost-constrained OFDM-based ISAC schemes in the asymptotic regime of large transmission blocks.
\subsection{Capacity}
We consider the asymptotic regime specified by $K \to \infty$ and $N \to \infty$, in which
achievable rates and capacity are defined in the standard fashion.
For fixed power and normalized ISL budgets $P$ and $S$, a rate $R$ is \emph{achievable} if there exists a sequence
of codebooks indexed by $(K,N)$, satisfying the power and ISL constraints
in \eqref{eq:power} and \eqref{eq:isl}, respectively, such that $\frac{1}{KN}\log M \to R$
and the maximal decoding error probability vanishes as $K,N \to \infty$.
The capacity $C(P,S)$ is the supremum of all such achievable rates.

Since the channel is memoryless and the constraints are on the empirical moments of the codewords, standard single-letterization arguments yield
\begin{equation}
\label{eq:C_PS_function}
\begin{aligned}
    C(P,S) \triangleq \ \sup_{F_X} \ I(X&;Y) \\
    \mathrm{s.t.} \ \ \E[|X|^2] &\leq P \\
    \kappa(X) - 1 &\leq S
\end{aligned}
\end{equation}
where $\kappa(X) \triangleq \E[|X|^4]/\big(\E[|X|^2]\big)^2$ is the kurtosis of the
input $X$. The second-moment constraint in \eqref{eq:C_PS_function} arises from the
per-codeword power constraint \eqref{eq:power}, and the kurtosis constraint from the
per-codeword ISL constraint \eqref{eq:isl}.

For the above capacity problem, the optimal input can be shown to exhaust the
power budget, since scaling any input to meet the power constraint with equality
strictly increases the mutual information while leaving the kurtosis unchanged. 
Therefore, the maximization in \eqref{eq:C_PS_function} can be taken over the set
\begin{equation}
    \mathcal{F}(P,S) \triangleq \{F_X : \E[|X|^2] = P, \ \E[|X|^4] \leq (1+S)P^2 \}.
    \label{eq:second_fourth_moment_constraints}
\end{equation}
\section{Optimal Signaling}
We now study optimal inputs that achieve capacity.
Since the moment constraints depend only on the input amplitude, standard arguments show that
capacity-achieving inputs can be taken to have phase uniformly distributed on
$(-\pi,\pi)$, independent of the amplitude \cite{Shamai1995}.
The amplitude distribution, in contrast, depends strongly on the ISL regime specified by $S$.
We begin with the two trade-off endpoints: the communication-optimal point, where the ISL constraint is inactive, and the sensing-optimal point,
where the ISL is zero.
\subsection{Communication-optimal point}
When the fourth-moment constraint in
\eqref{eq:second_fourth_moment_constraints} is inactive,
the capacity-achieving input distribution is Gaussian, i.e., $X_{\mathrm{G}} \sim \CN(0,P)$, for which $\E[|X_{\mathrm{G}}|^4] = 2P^2$ and $\kappa(X_{\mathrm{G}}) = 2$.
The ISL constraint is hence inactive when $S \geq 1$, and throughout this range the Gaussian input remains optimal, giving
\begin{equation}
    C(P,S) = \log \! \left( 1 + \frac{P}{\sigma^2} \right) \triangleq C_{\mathrm{G}}(P),
    \qquad S \geq 1.
\end{equation}
The communication-optimal endpoint is thus $C(P,1)$, where  $C_{\mathrm{G}}(P)$ is obtained with the lowest ISL budget.
\subsection{Sensing-optimal endpoint}
We now turn to the other extreme. The sensing-optimal point occurs at zero ISL, i.e., $S = 0$.
In this case, the fourth-moment constraint becomes $\E[|X|^4] \leq P^2$. On the other hand, Jensen's inequality gives $\E[|X|^4] \geq \big(\E[|X|^2]\big)^2 = P^2$, with equality if and only if $|X|$ is constant almost surely.
Combining the two, we conclude that input $X$ must be constant modulus for $S = 0$. 

Under such constraint, it can be shown that the optimal input is $X_{\mathrm{CM}} \triangleq \sqrt{P}\, e^{\imag \Theta}$, with $\Theta \sim \mathcal{U}(-\pi,\pi)$ being a uniform phase. 
The corresponding sensing-optimal capacity is \cite{Wyner1966}
\begin{equation}
    C(P,0) = \frac{2P}{\sigma^2} - \E \bigg[ \log I_0 \bigg( \frac{2\sqrt{P}}{\sigma^2}|Y| \bigg) \bigg] \triangleq C_{\mathrm{CM}}(P)
\end{equation}
where $I_0(\cdot)$ is the modified Bessel function of the first kind of order zero, and $|Y|$ is Rician distributed with
\begin{equation}
    f_{|Y|}(w) = \frac{2w}{\sigma^2} \exp \! \left( - \frac{w^2 + P}{\sigma^2} \right) I_0 \bigg( \frac{2 \sqrt{P}\, w }{\sigma^2} \bigg).
\end{equation}
\subsection{Intermediate regime}
The intermediate regime is characterized by $0 < S < 1$. Here the fourth-moment
constraint in \eqref{eq:second_fourth_moment_constraints} can be shown to be
active, and the optimal input distribution has no known closed form.
Nevertheless, its structure can be inferred from the work of Gursoy \emph{et al.}
\cite{Gursoy2005}, who studied the capacity-achieving input of the Rician channel
under second- and fourth-moment constraints.
In particular, \cite[Theorem~1]{Gursoy2005} shows that whenever the fourth-moment
constraint is active, the capacity-achieving amplitude is discrete with finitely
many mass points.
Specializing this argument to our channel, the capacity-achieving input in the
intermediate regime $0<S<1$ inherits the same discrete, finite-support amplitude structure,
yielding a concentric multi-ring geometry.
\begin{remark}
Recent constellation-shaping studies for ISAC have empirically found
APSK geometries to perform well in the
intermediate regime \cite{Shao2026, Keshavarzchafjiri2026}. Our characterization
provides a capacity-theoretic justification for this observation, as the
capacity-achieving input consists of concentric rings whose natural
finite-cardinality realization is an APSK constellation.
At the sensing-optimal endpoint, the rings collapse to a single radius with
continuous uniform phase, whose finite-cardinality counterpart is phase-shift keying (PSK). This is consistent with PSK having been argued to be the sensing-optimal
constellation \cite{Du2024, FLiu2025a, Geiger2026}.
\end{remark}

\section{Superposition Signaling}
\label{sec:method}
We now propose a new signaling strategy, based on scaling and superimposing the
communication-optimal and sensing-optimal inputs identified in the previous
section.
As we shall see, the proposed strategy is easy to analyze, and furthermore strikes a trade-off between communication and sensing by tuning a single power-allocation
parameter.

Given a power allocation parameter $\lambda \in [0,1]$, the single-letter superposition input is given by
\begin{equation}
    X_{\lambda} = \sqrt{1-\lambda} \, X_{\mathrm{G}} + \sqrt{\lambda} \, X_{\mathrm{CM}} \label{eq:superposition}
\end{equation}
where $X_{\mathrm{G}} \sim \CN(0,P)$ is the communication-optimal Gaussian input
and $X_{\mathrm{CM}} = \sqrt{P}\, e^{\imag \Theta}$, with
$\Theta \sim \mathcal{U}(-\pi,\pi)$, is the sensing-optimal constant-modulus input.
Note that $X_{\mathrm{G}}$ and $\Theta$ (and hence $X_{\mathrm{CM}}$) are taken to
be statistically independent.
The second and fourth moments of $X_{\lambda}$ are
\begin{align}
    \E[|X_{\lambda}|^2] &= P, \\
    \label{eq:fourth-moment_SP}
    \E[|X_{\lambda}|^4] &= (2-\lambda^2)\,P^2.
\end{align}
As a sanity check, $\lambda = 0$ recovers the Gaussian-input with fourth moment $2P^2$, while
$\lambda = 1$ recovers the constant-modulus input with fourth moment $P^2$.
\subsection{Achievable rate and successive decoding}
Using standard achievability arguments, the rate achieved by superposition
signaling is $I(X_{\lambda};Y)$. This can be factorized as
\begin{align}
    I(X_{\lambda};Y)
    &= I(X_{\lambda}, X_{\mathrm{CM}} ; Y) \label{eq:spmi0} \\
    &= I(X_{\mathrm{CM}};Y) + I(X_{\lambda};Y \mid X_{\mathrm{CM}}) \label{eq:spmi1} \\
    &= I(X_{\mathrm{CM}}; Y) + I(X_{\mathrm{G}}; \sqrt{1-\lambda} \, X_{\mathrm{G}} + Z) \label{eq:spmi2}
\end{align}
where \eqref{eq:spmi0} holds due to the Markov chain
$X_{\mathrm{CM}} \to X_{\lambda} \to Y$.
The term $I(X_{\mathrm{CM}};Y)$ is exactly the capacity under constant-modulus signaling with an SNR of $\lambda P / ((1-\lambda) P + \sigma^2)$, resulting from treating the Gaussian input component as additional noise.
On the other hand, $I(X_{\mathrm{G}};\sqrt{1-\lambda} \, X_{\mathrm{G}}+Z)$ is the capacity under Gaussian signaling at a reduced SNR of $(1-\lambda)P/\sigma^2$.
Therefore, we can express the achievable rate under superposition signaling as
\begin{equation}
    I(X_{\lambda};Y)
    = C_{\mathrm{CM}} \! \left( \frac{\lambda P}{(1-\lambda) P/\sigma^2+1} \right) + C_{\mathrm{G}} ( (1-\lambda) P ) \label{eq:sp_rate}
\end{equation}
which clearly represents the contributions of each component in the superposition input \eqref{eq:superposition}.

The above decomposition admits a natural interpretation in terms of superposition
coding and successive decoding.
In particular, a scheme can be constructed by independently generating a
sensing-optimal codebook at a rate up to $I(X_{\mathrm{CM}}; Y)$ and a
communication-optimal codebook at a rate up to
$I(X_{\mathrm{G}};\sqrt{1-\lambda} \, X_{\mathrm{G}}+Z)$.
Each transmitted codeword is the superposition of one codeword from each codebook.
The decoder performs successive decoding, first recovering the constant-modulus
codeword while treating the Gaussian codeword as noise, then decoding the Gaussian
codeword given the constant-modulus codeword.
Note that this incurs no loss in rate performance compared to joint decoding.
\subsection{Superposition rate}
For fixed $P$ and $S$, we define the superposition rate $C_{\mathrm{SP}}(P,S)$
as the maximum achievable rate $I(X_{\lambda};Y)$ over $\lambda \in [0,1]$ subject to
$\E[|X_{\lambda}|^4] \leq (1+S)P^{2}$.
From \eqref{eq:fourth-moment_SP}, this constraint is equivalent to
$\lambda \geq \sqrt{1-S}$. Since $I(X_{\lambda};Y)$ is decreasing in $\lambda$, it
is maximized at $\lambda = \sqrt{1-S}$, giving
\begin{equation}
    C_{\mathrm{SP}}(P,S) = I(X_{\sqrt{1-S}};Y).
\end{equation}
It is clear that $C_{\mathrm{SP}}(P,S) = C(P,S)$ at the endpoints $S = 0$ and $S = 1$.
However, in the intermediate regime $0 < S < 1$, it is not clear a priori how much
$C_{\mathrm{SP}}(P,S)$ falls short of $C(P,S)$. In terms of input distribution,
the superposition scheme is circularly symmetric but has a continuous amplitude profile, unlike the optimal input, whose amplitude is discrete.
Nevertheless, the gap turns out to be very small, as illustrated by the numerical results in the following section. 
\section{Bounds and Numerical Results}
We now present bounds on the capacity $C(P,S)$ that will be used to benchmark the
proposed superposition signaling scheme.
These bounds were originally derived by Geiger \emph{et al.} \cite{Geiger2026}, and we re-derive
them here for completeness.
\subsection{Capacity upper bound}
The upper bound in  \cite{Geiger2026} is obtained by maximizing the output differential entropy over a relaxed set of distributions
\begin{equation}
    C(P,S) \leq \sup_{F_Y \in \mathcal{F}_Y(P,S)} h(Y) - h(Z) \label{eq:cap_UB}
\end{equation}
where $h(Z) = \log(\pi e \sigma^2)$ and the constraint set is given by
\begin{equation}
\begin{aligned}
\mathcal{F}_Y(P,S) \triangleq  \{& F_Y :  \E[|Y|^2] =  P + \sigma^2,\\
& \E[|Y|^4] \leq (1+S)P^2 + 4 P \sigma^2 + 2 \sigma^4 \}.
\end{aligned}
    \label{eq:output_moment_constraints}
\end{equation}
The constraints in \eqref{eq:output_moment_constraints} are determined by the
combined effect of the input constraints in
\eqref{eq:second_fourth_moment_constraints} and the noise, using the independence
and circular symmetry of $X$ and $Z$ for the fourth moment. The capacity upper bound holds since every output
induced by a feasible input satisfies the constraints in \eqref{eq:output_moment_constraints}.

Let $\tilde{Y}$ be the output attaining \eqref{eq:cap_UB}. From the maximum entropy principle \cite[Ch.~12]{Cover2006}, its density is given by 
\begin{equation}
    f_{\tilde{Y}}(y) = \tfrac{1}{\gamma} e^{-\alpha |y|^2 - \beta |y|^4 } \label{eq:max_output}
\end{equation}
where $\gamma \geq 0$, $\beta \geq 0$, $\alpha \in \mathbb{R}$ are such that $\tilde{Y}$ satisfies the constraints in \eqref{eq:output_moment_constraints}.
The resulting upper bound is given by 
\begin{equation}
    C_{\mathrm{ub}}(P,S) \triangleq h(\tilde{Y}) - h(Z) \label{eq:ub}
\end{equation}
where
\begin{align}
h(\tilde{Y}) 
&= \log \gamma + \alpha \, \E[|\tilde{Y}|^2] + \beta \, \E[|\tilde{Y}|^4] \\
&= \log \gamma + \alpha (P + \sigma^2) + \beta ((1+S)P^2 + 4P\sigma^2 + 2\sigma^4). \notag
\end{align}
Note that the fourth-moment constraint on the output can be shown to hold with
equality at the optimum.

The distribution parameters $\gamma$, $\alpha$, and $\beta$ can be found using the second and fourth moments of the distribution.
Due to the coupling among the parameters \cite{Geiger2026}, we have
\begin{align}
    \alpha \, \E[|\tilde{Y}|^2] + 2 \beta \, \E[|\tilde{Y}|^4] &= 1, \label{eq:coupling1} \\
    \alpha + 2 \beta \, \E[|\tilde{Y}|^2] &= \pi/\gamma. \label{eq:coupling2}
\end{align}
Using \eqref{eq:coupling1}, $\beta$ can be fully determined from $\alpha$ such that
\begin{equation}
    \beta(\alpha) = \frac{1- \alpha \E[|\tilde{Y}|^2]}{2 \, \E[|\tilde{Y}|^4]} \label{eq:beta}.
\end{equation}
Moreover, using \eqref{eq:coupling2}, $\gamma$ is written in terms of $\alpha$ as
\begin{equation}
    \gamma(\alpha) = \frac{\pi}{\alpha + 2 \, \beta(\alpha) \E[|\tilde{Y}|^2]}. \label{eq:gamma}
\end{equation}
Hence, the maximum-entropy distribution in \eqref{eq:max_output} can be controlled by a single parameter $\alpha$.

Finally, to satisfy $\int f_{\tilde{Y}}(y) \, d y = 1$, we must have
\begin{equation}
    \gamma(\alpha) = \sqrt{\frac{\pi^3}{\beta(\alpha)}} \exp \! \left( \frac{\alpha^2}{4\beta(\alpha)} \right) Q \! \left( \frac{\alpha}{\sqrt{2\beta(\alpha)}} \right) \label{eq:scaling}
\end{equation}
where $Q(x) \triangleq \frac{1}{\sqrt{2\pi}} \int_x^{\infty} \exp(-t^2/2) \, dt$ is the $Q$-function.
$\alpha$ can be found numerically by equating \eqref{eq:gamma} and \eqref{eq:scaling}.
\subsection{Capacity lower bound}
A lower bound is also obtained in \cite{Geiger2026} using the entropy power
inequality \cite[Ch.~17.8]{Cover2006}, given by
\begin{equation}
    C(P,S) \geq C_{\mathrm{lb}}(P,S) \triangleq \log \! \big( e^{h(\tilde{X})} + e^{h(Z)} \big) - h(Z)
    \label{eq:lb}
\end{equation}
where $\tilde{X}$ maximizes the differential entropy subject to the input moment
constraints in \eqref{eq:second_fourth_moment_constraints}. This maximization is
carried out in almost the same manner as the output-entropy maximization used above to
derive the upper bound.

\begin{figure*}[t]
    \centering
    \includegraphics[width=0.98\textwidth]{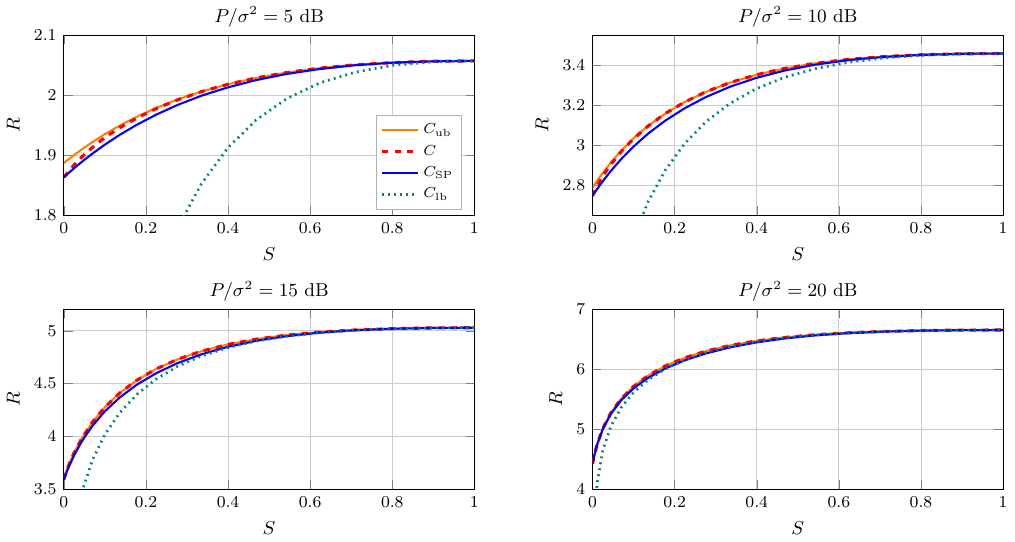}
    \vspace{-2.5mm}
    \caption{The performance of the superposition scheme against the benchmarks.}
    \label{fig:curves}
\end{figure*}

\subsection{Numerical results}
\label{sec:results}
\begin{figure}[t]
    \centering
    \includegraphics[width=\columnwidth]{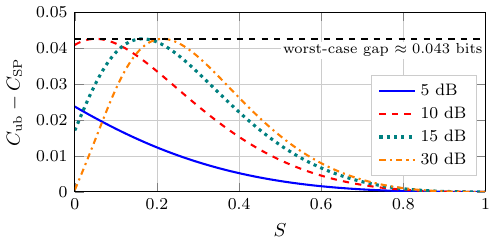}
    \vspace{-7mm}
    \caption{The rate gap between the upper bound and superposition signaling.}
    \label{fig:gap}
\end{figure}
\begin{figure*}[t]
    \centering
    \includegraphics[width=\linewidth]{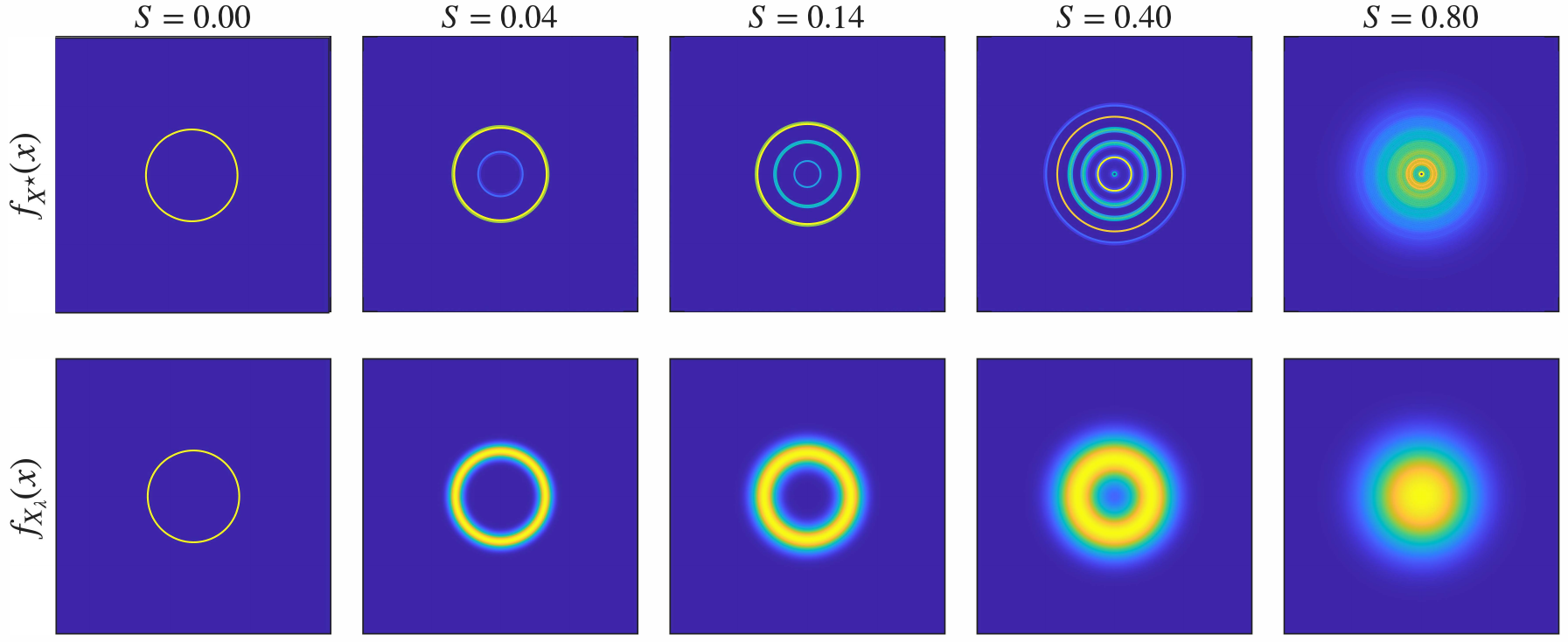}
    \vspace{-5mm}
    \caption{Illustration of the distributions of the optimized input $f_{X^{\star}}(x)$ and the superposition input $f_{X_{\lambda}}(x)$ for $P/\sigma^2= 10$ dB.}
    \label{fig:densities}
\end{figure*}
We now evaluate the performance of the superposition signaling scheme.
In Fig.~\ref{fig:curves}, we plot the bounds on the capacity and the rate under superposition signaling against the ISL budget for different SNR values. All rates are expressed in bits per complex channel use.
We also include a numerical approximation of the capacity itself, computed via a discretized optimization procedure (see the Appendix for details).

As can be seen, the upper bound is close to capacity across all scenarios.
Moreover, the superposition signaling rate is close to both the capacity and its upper bound, with a small gap in the intermediate regime.
The entropy power inequality lower bound in \eqref{eq:lb}, on the other hand, is loose in general. It becomes tight only as $S$ increases and the optimal input approaches the Gaussian distribution, or as the additive noise becomes negligible.
Note that the superposition rate $C_{\mathrm{SP}}(P,S)$ can be viewed as an
alternative lower bound on the capacity, which is both tight and easy to evaluate.

Fig.~\ref{fig:gap} plots the gap between the upper bound and the superposition rate, in bits, for different SNR values.
Note that this gap also upper-bounds the rate gap between optimal and superposition signaling.
We observe that the gap is at most $\approx 0.043$ bits over all considered
scenarios, which is remarkably small given that the superposition input has
continuous amplitude support, unlike the capacity-achieving input.

Finally, in Fig.~\ref{fig:densities}, we illustrate the optimized input distribution using discretization, which is denoted by $f_{X^{\star}}(x)$, and the superposition input density $f_{X_{\lambda}}(x)$.
Both distributions start with a single ring at $S=0$, i.e., the sensing-optimal point.
As $S$ increases, the number of rings in the optimized input increases and the superposition input allocates more power to the Gaussian input.
As $S \to 1$, both distributions start to look closer to a Gaussian, as expected.
\section{Conclusion}
\label{sec:conclusion}
We studied the capacity of the complex AWGN channel when the input is subject to second-moment and kurtosis constraints.
This models an OFDM-ISAC system when the ISL is used as the sensing cost.
The capacity-achieving signaling strategy at the sensing-optimal and communication-optimal endpoints was characterized by circularly symmetric constant-modulus signaling and complex Gaussian signaling, respectively.
We proposed the superposition of these two inputs as a viable signaling strategy that can be easily tuned to meet various trade-offs.
We demonstrated the performance of the proposed scheme via numerical results, and showed that the worst-case gap to the capacity is small.

Providing theoretical guarantees on the performance of the superposition scheme is left for future work.
Moreover, extending the analysis and design strategies to more general settings, such as frequency-selective communication channels, constitutes another promising research direction.
\appendix
In this part, we summarize the numerical procedure to compute the capacity.
We set the input $X$ to be circularly symmetric, which yields a circularly symmetric output $Y$.
Define $R \triangleq |X|$ and $W \triangleq |Y|$.
Then, we can write \cite{Shamai1995}
\begin{equation}
    I(X;Y) = h(W) + \E[\log W] - 1 + \log \! \left(\tfrac{2}{\sigma^2} \right). \label{eq:mi}
\end{equation}
We select $N_r =100$ and $N_w=10000$ uniformly spaced input and output amplitude mass points, respectively, such that
\begin{equation}
    r_i \triangleq \frac{i-1}{N_r-1} r_{\max} \quad \text{and} \quad w_j \triangleq \frac{j-1}{N_w-1} w_{\max}
\end{equation}
where we set $r_{\max} = 4 \sqrt{P}$, and $w_{\max} = r_{\max} + 8 \sigma$.
We define $p_R(r_i) \triangleq \Prb[R=r_i]$ as the probability mass function (pmf) of the input amplitude. Then, the conditional pmf of the output amplitude given the input amplitude is defined as
\begin{equation}
    p_{W|R}(w_j | r_i) \triangleq \frac{ f_{W|R}(w_j|r_i)}{\sum_{k=1}^{N_w} f_{W|R}(w_k|r_i) }.
\end{equation} 
Here, $W|R=r$ follows a Rician distribution such that
\begin{equation}
    f_{W|R}(w|r)
    = \frac{2 w}{\sigma^2} \, \exp \! \left( -\frac{w^2 + r^2}{\sigma^2} \right) I_0 \! \left( \frac{2 w r}{\sigma^2} \right). \label{eq:channel_law}
\end{equation}
Then, the pmf of the output amplitude induced by an input amplitude with pmf $p_R$ is given by
\begin{equation}
    p_W(w_j;p_R) \triangleq \sum_{i=1}^{N_r} p_R(r_i) \, p_{W|R}(w_j | r_i).
\end{equation}
Finally, set $\Delta w \triangleq w_{\max}/(N_w-1)$; using \eqref{eq:mi}, the capacity is approximated by a concave optimization problem given by
\begin{align}
C(P,S)& \approx \max_{p_R} \;
  \sum_{j=1}^{N_w} p_W(w_j;p_R) \log\frac{w_j}{p_W(w_j;p_R)}
    \!+\! \log\!\frac{2\Delta w}{e\sigma^2} \notag \\
\text{s.t.}\quad
  & p_R(r_i) \geq 0,
  && \hspace{-5cm} \sum_{i=1}^{N_r} p_R(r_i) = 1, \label{eq:opt} \\
  & \sum_{i=1}^{N_r} r_i^2\, p_R(r_i) = P,
  && \hspace{-5cm} \sum_{i=1}^{N_r} r_i^4\, p_R(r_i) \leq (1+S)\,P^2. \notag
\end{align}
\balance
\bibliographystyle{IEEEtran}
\bibliography{bibliography}
\end{document}